\documentclass[conference]{IEEEtran}
\IEEEoverridecommandlockouts
\usepackage{cite}
\usepackage{amsmath,amssymb,amsfonts}
\usepackage{algorithmic,hyperref}
\usepackage{graphicx}
\usepackage{textcomp}
\usepackage{xcolor}
\def\BibTeX{{\rm B\kern-.05em{\sc i\kern-.025em b}\kern-.08em
    T\kern-.1667em\lower.7ex\hbox{E}\kern-.125emX}}

\newtheorem{lemma}{Lemma}

\newtheorem{corollary}{Corollary}
    
\newcommand{\OO}{\ensuremath{\mathcal O}}
    
\begin{document}

\title{Towards Privacy in Geographic Message Dissemination for Connected Vehicles
\thanks{A revised version of this paper was published in 2019 IEEE International Conference on Connected Vehicles and Expo (ICCVE), \url{https://doi.org/10.1109/ICCVE45908.2019.8965198}.}
}

\author{\IEEEauthorblockN{Stefan Ruehrup}
\IEEEauthorblockA{ASFINAG \\
Telematic Services \\
Vienna, Austria \\
stefan.ruehrup@asfinag.at}
\and
\IEEEauthorblockN{Stephan Krenn}
\IEEEauthorblockA{AIT Austrian Institute for Technology \\
Digital Safety \& Security \\
Vienna, Austria \\
stephan.krenn@ait.ac.at}
}

\maketitle

\begin{abstract} 
  With geographic message dissemination, connected vehicles can be served with traffic information in their proximity, thereby positively impacting road safety, traffic management, or routing.
  Since such messages are typically relevant in a small geographic area, servers only distribute messages to affected vehicles for efficiency reasons.
  One main challenge is to maintain scalability of the server infrastructure when collecting location updates from vehicles and determining the relevant group of vehicles when messages are distributed to a geographic relevance area, while at the same time respecting the individual user's privacy in accordance with legal regulations.

  In this paper, we present a framework for geographic message dissemination following the privacy-by-design and privacy-by-default principles, without having to accept efficiency drawbacks compared to conventional server-client based approaches.
\end{abstract}

\begin{IEEEkeywords}
connected vehicles, consistent hashing, space-filling curves, privacy-by-design
\end{IEEEkeywords}

\section{Introduction}

There are several use cases for connected vehicles ranging from distribution of traffic information to the exchange of real-time event and object data to support advanced assistance functions and automated driving functions. A basic use case -- and a very important one from traffic safety point of view -- is the distribution of traffic information and event messages to connected vehicles, where the information may be generated in the traffic management center (TMC) or by other vehicles. 

\begin{figure}[htbp]
\centerline{\includegraphics[width=.75\linewidth]{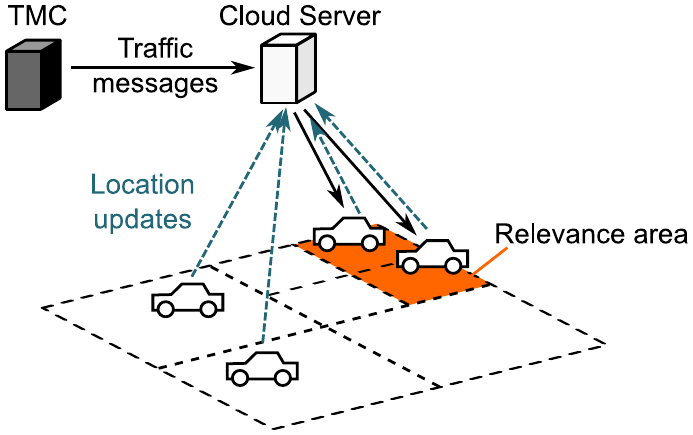}}
\caption{Conventional client/server model for message distribution}
\label{fig:conventional-model}
\end{figure}

The vast majority of traffic messages are related to a certain road segment, a location or an area, thus there is a relevance area connected to a message. Consequently such messages should be forwarded to the vehicles within this relevance area, and not necessarily to all vehicles. A typical implementation to serve such use case is a client-server structure, where vehicles regularly announce their geographic location to the server and the server, in turn, distributes messages relevant for these clients. Hence, relevance is based on the proximity between a vehicles' location and the location reference of the message, cf. also Fig.~\ref{fig:conventional-model}. It is possible that the server is requested for traffic information (pull) rather than pushing messages to subscribed clients. However, a publish subscribe mechanism is more suitable for near real-time messaging. There are several message queues and protocols that serve this goal, such as AMQP or MQTT. 

As in many location based services, each client needs to report its location to a server in order to receive messages that are relevant for its location. Thus the mobility of clients could be tracked by the central server, posing a potentially severe privacy risk, also in light of recent legal regulations~\cite{gdpr}.

This leads to the following observations:

\begin{itemize}
\item Client-centric subscription: A client (vehicle) should subscribe to its geographic area of interest rather than being tracked by a server. This way home location, movement trajectories and destination location etc. can be covered based on client decisions instead of reporting and tracking the actual position.
\item Variable location reporting: In order to maximally benefit from the above, the client should be able to arbitrarily vary the precision of reported location.
\item Message Routing (Queuing and forwarding): Since the areas of relevance are known to the sender, messages should be routed by a network of routers instead of server-based translation or transposition.
\end{itemize}

In order to fulfill the aforementioned requirements, we propose a routing network that forwards messages to connected vehicles, where the assignment of vehicles to routers is based on distributed hash functions. All geolocations are encoded on a Z-order curve, a 1-dimensional space filling curve that preserves proximity and allows efficient range queries. Z-order locations have the property that the length of the prefix determines location accuracy (the shorter the bit-string the larger the region around the actual position).

\begin{figure}[htbp]
\centerline{\includegraphics[width=.8\linewidth]{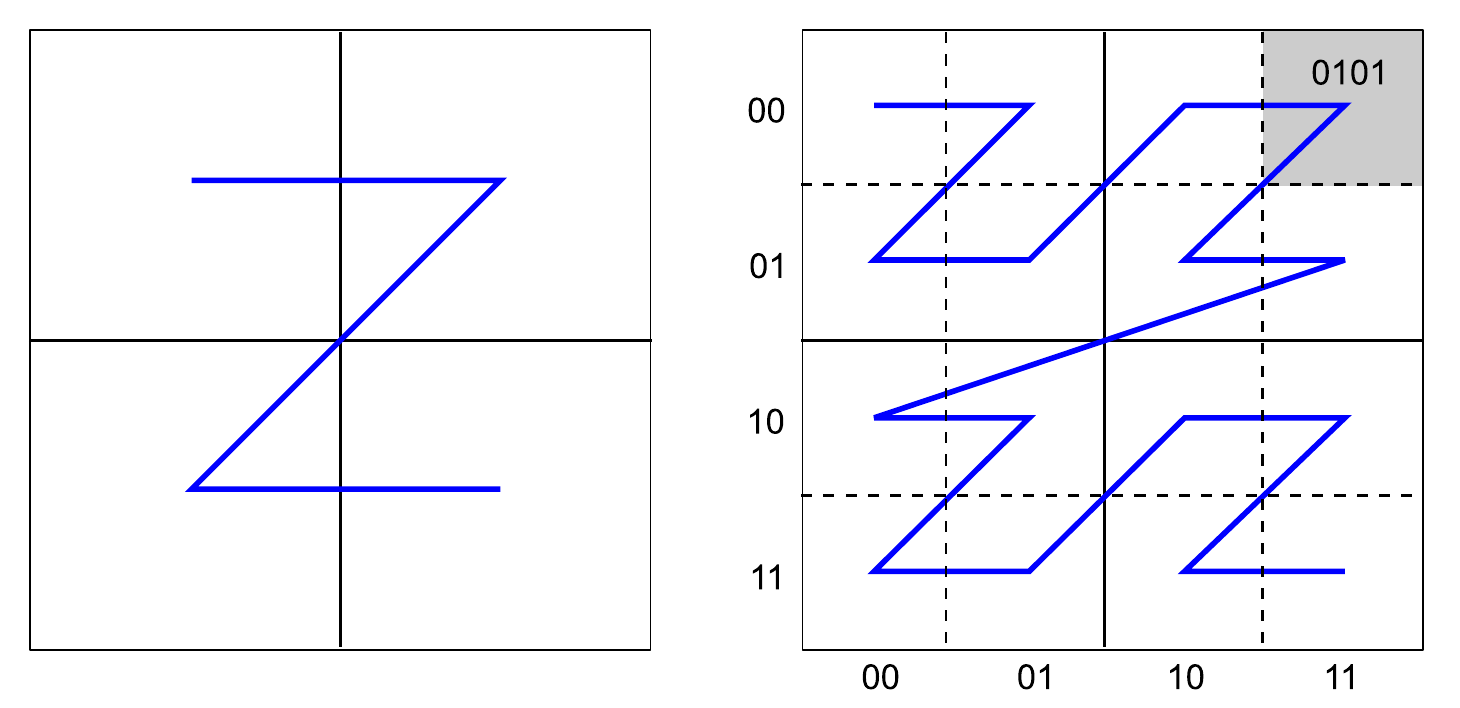}}
\caption{Z-order curve: for x-coordinate $X=x_0x_1$ (in binary) and y-coordinate $Y=y_0y_1$, a specific area has index $y_0x_0y_1x_1$, thereby guaranteeing that areas with identical prefix are close to each other.}
\label{fig:z-curve}
\end{figure}

\section{Related Work}

Mapping of geocoordinates to 1-dimensional space filling curves has been well studied in the literature. The Z-order, which we consider in this paper, has been studied for indexing data structures for efficient range searching in  \cite{tropf81multidimensional} and \cite{orenstein84class}, cf. also Fig.~\ref{fig:z-curve}. An overview of results on range searching is given in \cite{agarwal17range}. Proximity has been early identified as a criterion for efficient range queries. Asano et al. show that specific performance metrics could be optimised by a proper choice or adaption of the space filling curve \cite{asano97space}. Lower bounds on the proximity for the Z curve are presented in \cite{xu12lower}. We use the space filling curve also to define the assignment between clients and routers. The consistent hashing scheme \cite{karger97consistent} is well known for providing an assignment between clients and servers in a distributed way using a common hash function. Consistent hashing has led to the definition of distributed hash tables (DHTs) in the area of peer-to-peer (P2P) networks, especially Chord \cite{stoica01chord} in order to maintain a routing network and provide lookups in a distributed way (see \cite{mahlmann07peer,lua05survey} for an overview).

Privacy in location-based services has been extensively studied in the academic literature, e.g., \cite{gedlui05,meycho09,hgxa07,lykkl16}. 
The presented approaches include waiting for at minimum of $k$ users before serving them~\cite{gedlui05}, which however delays realtime operations in sparse scenarios or regions.
Alternatively, a trusted proxy prefetching information on behalf of a user was suggested~\cite{hgxa07}, which however introduces certain undesirable trust assumptions.
Finally, obfuscating users' paths by generating fake paths was suggested~\cite{lykkl16}.
This approach requires inter-vehicle communication and frequent updates of network addresses, and assumes that the density of cars is not too sparse, as it relies on mutually obfuscating movement patterns. 
Furthermore, this approach harms functionality as the generated fake paths no longer allow proper traffic load analysis or similar.
A good overview is given in the recent work of Asuquo et al.~\cite{acmolhbs18}.
In this work we propose some pragmatic approaches which give the user tunable privacy guarantees without harming the information that an infrastructure provider requires.

\section{System Model}

\begin{figure}[htbp]
\centerline{\includegraphics[width=\linewidth]{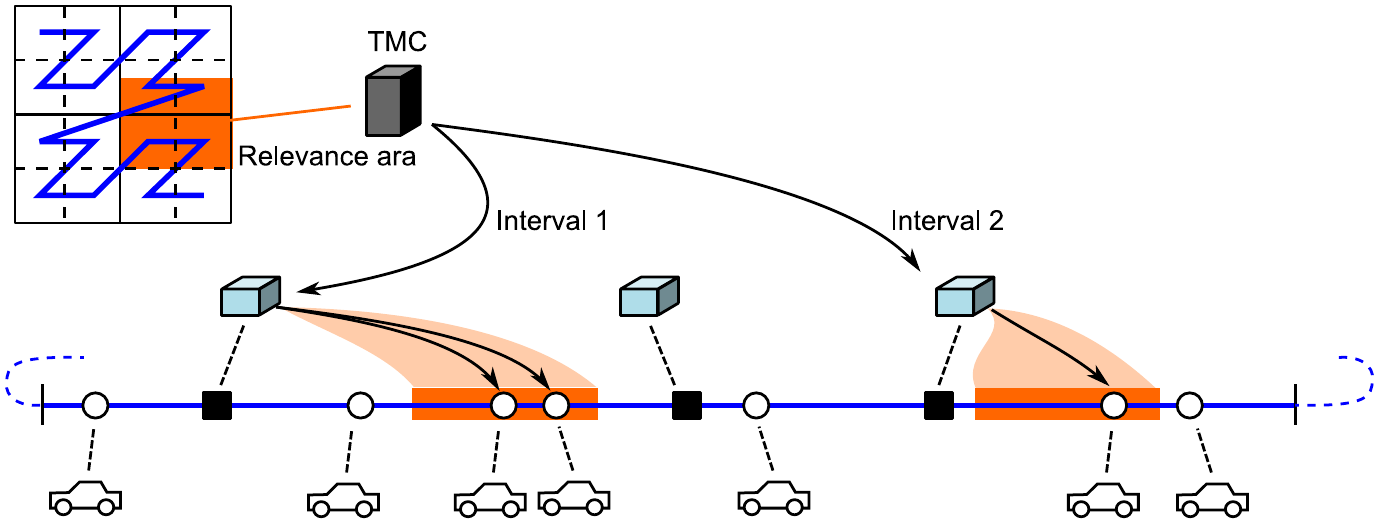}}
\caption{Message routing model: A message generated by the TMC is forwarded to router(s) responsible for the relevance area. These routers forward the message to connected vehicles.}
\label{fig}
\end{figure}

We consider the following elements in our system model:

\subsection{Clients} 
A client (vehicle) connects to the network by contacting a known routing node, from which it learns other routers in the network. All routers in the network have one or more virtual locations. A moving vehicle (active client) assigns itself to a router based on its own real location and the router's virtual location. Location changes of the vehicle are reported to the assigned router, especially if a location change constitutes a change of the assigned router. The trigger condition for location updates and the precision of location information depends on the privacy preferences of the vehicle (cf.\ Section~\ref{sec:dynamic-region-granularity}). 

\subsection{Routers} 
A router distributes messages to clients. A router has one or more virtual locations. The virtual locations determine the geographic regions for which the router is responsible. All clients within the responsible region connect to the router. The router maintains a routing table of all clients, using a data structure that uses the Z-order as sorting key and for performing range queries. Routers exchange information about their virtual locations with other routers. Depending on the overall size of the routing network, a fully meshed network of routers or an overlay network such as Chord \cite{stoica01chord} could be used (cf.\ Section \ref{sec:alternative-routing-network}).

\subsection{Locations} 
The Z-order curve (Fig.~\ref{fig:z-curve}) is used for a one-dimensional representation of two-dimensional geo-coordinates. This representation can be treated as a ring when assuming wrap-around ends. Client locations and virtual router locations are points on the ring and form the assignment. Virtual router locations are initially randomly distributed and can be re-allocated for load-balancing reasons during runtime. 

Note that in contrast to consistent hashing \cite{karger97consistent} we do not use purely random assignments. While clients (vehicles) use their real location, routers use a set of random virtual locations following an empirical distribution function of the vehicles. 

Note further that other space-filling curves such as the Hilbert curve also preserve proximity and can be used as alternative to the Z-order curve.

\begin{figure}[htbp]
\centerline{\includegraphics[width=\linewidth]{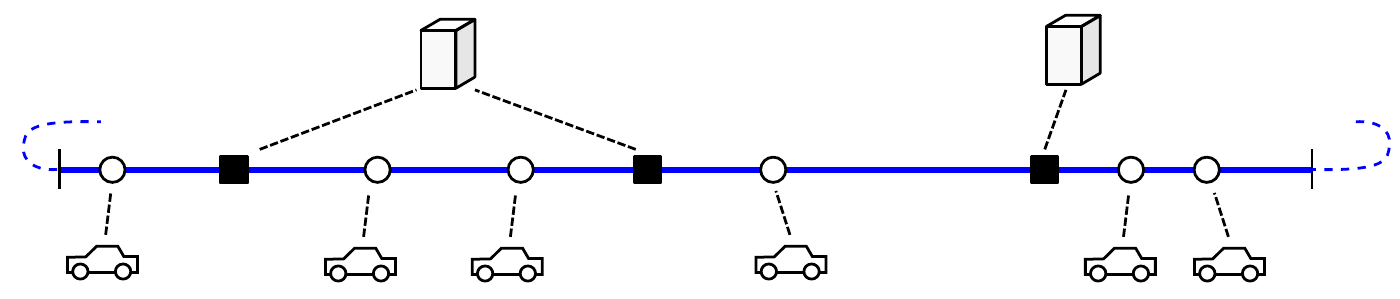}}
\caption{Distributed hashing scheme for the assignment between routers ($\blacksquare$) and vehicles ($\bigcirc$). Each vehicle is assigned to the router having a virtual location in descending Z-order, i.e. to the left in the figure)}
\label{fig:consistent-hashing}
\end{figure}

\section{Privacy Aspects}
In the following we present potential approaches that could be used in our context in order to guarantee privacy-by-design and privacy-by-default, and briefly describe pros and cons of each of these approaches.
Note that these approaches are not mutually exclusive and could easily be combined with each other to achieve the best privacy results.

\subsection{Polling vs. Publish/Subscribe}
  In a publish/subscribe based setting, a vehicle needs to register itself at a server every time it enters an area for which it wishes to receive updates, which allows the server to track a vehicle.
  Changing TCP connections or even the network address of the vehicle every time when entering a new sector or also on a regular basis does not solve the problem either, as it still allows for statistical correlations between expires/closed and newly established connections.

  A potential approach is to switch to a polling based system, where the vehicles periodically fetch the available information of their sectors of interest, using different pseudonyms and connections for every such access.
  For instance using attribute-based credential systems, first envisioned by Chaum~\cite{chaum85}, the vehicles could pseudonymously prove to the server that they are genuine cars, thereby mitigating denial of service attacks through other network participants.
  Furthermore, such systems give the user fine-granular control under which conditions different pseudonyms should be linkable and when they should be fully independent from each other, such that average travel times or speeds could still be made available to the infrastructure provider.
  We refer the interested reader to~\cite{cklmnp15} for details on the cryptographic modeling, and to~\cite{klss17} for a cloud-based instantiation of such credential systems in order to minimize the computational power needed within the vehicle's on-board unit.

  While such a polling-based approach gives high privacy guarantees, it increases the bandwidth requirements and increases the computational load especially on the server, as clients need to pull with high frequency in order to support (near) realtime updates in case of, e.g., danger spots.

\subsection{Layered Routing}
\begin{figure}[htbp]
    \centerline{\includegraphics[width=\linewidth]{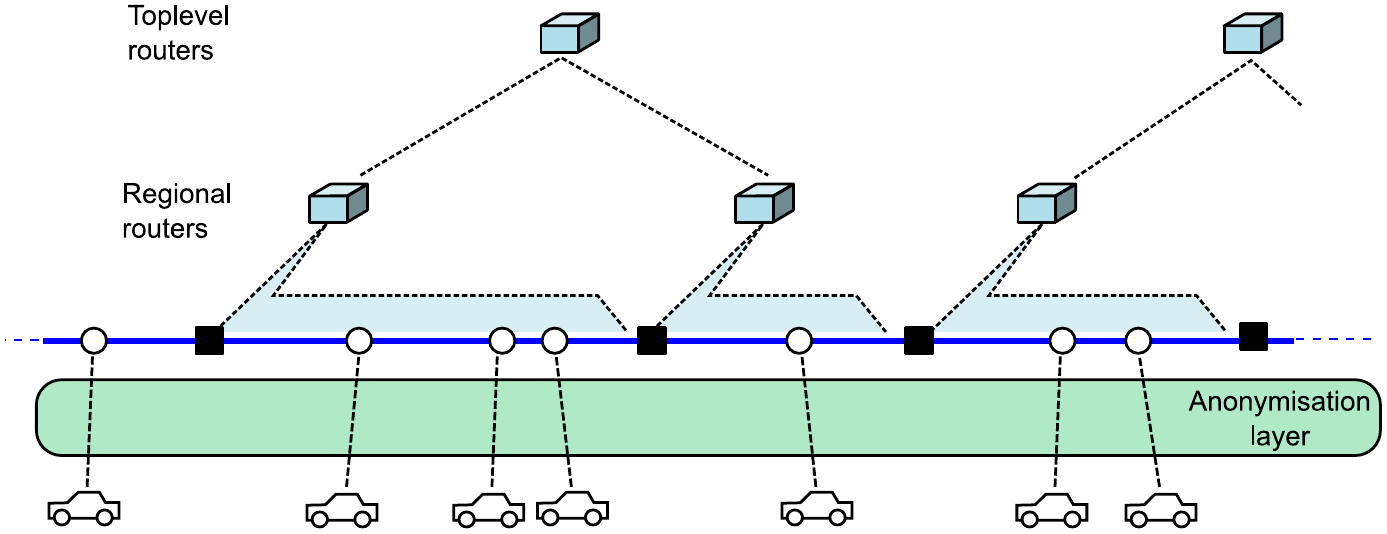}}
    \caption{Layered Routing}
    \label{fig:layered_routing}
  \end{figure}

  A layered routing approach could be used to avoid a single entity from receiving all location information of a client. 
  In this setting, a top layer of servers is responsible for routing in large areas while vehicles connect only to regional routers that are responsible for a specific area only, cf. also Fig.~\ref{fig:layered_routing}. 
  Such a setup may be realistic in scenarios where different parts of the road network are operated by different entities, such as different federate states or counties.

  Using this approach allows the regional routers to locally track vehicles, without giving them higher level information in case of long-distance trips.
  On the other hand, the top-level routers only learn very coarse information about a specific vehicle, but detailed origin-destination relations are disguised.

  This way, message dissemination to large areas or cross-border areas is possible while hiding vehicle links and IP addresses from a few entities. 
  A further anonymisation layer, e.g. TOR, could be added to hide IP addresses from the routing network.

\subsection{Dynamic Region Granularity} \label{sec:dynamic-region-granularity}

  In order to preserve the privacy of the individual user without at the same time undermining the functionality of the system, the granularity for which a vehicle requests location information may vary depending on different parameters, such as time or vehicle density.
  By doing so, $k$-anonymity can be achieved, i.e., it can be ensured that at every point in time the vehicle is anonymous within a set of at least $k$ vehicles.
  Meaningful values for $k$ may again depend on various parameters, such as average number of rides, or how often vehicle cross region boundaries, and determining specific values is beyond the scope of this work.

\begin{figure}[htbp]
    \centerline{\includegraphics[width=.6\linewidth]{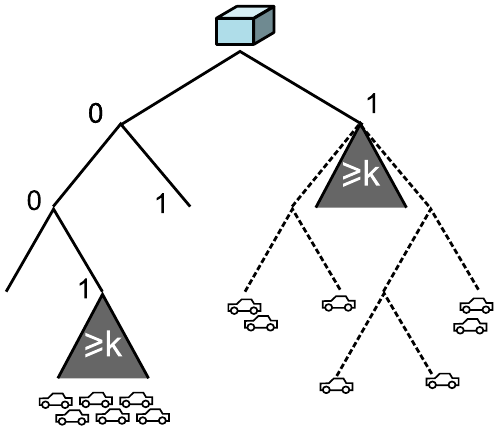}}
    \caption{$k$-anonymity: For each prefix it should be ensured that the number of vehicles that can be addressed is higher than a given $k$. This is fulfilled in the left branch for $k=5$ and prefix $001$ and fulfilled in the right branch for prefix $1$.}
    \label{fig:k-anonymity}
  \end{figure}

  More specifically, in our system areas with matching prefixes are geographically close to each other.
  Thus, by truncating the last bits of the location when contacting a router, vehicles can dynamically decide how detailed location information they reveal to the infrastructure provider, cf. also Fig.~\ref{fig:k-anonymity}.

  This can now be leveraged in two ways:

  \begin{enumerate}
    \item
      Firstly, following the privacy-by-design and privacy-by-default principles, routers can announce the maximum length of the prefix that vehicles should report in order to guarantee that at least $k$ vehicles are within the same such area, depending on the region, traffic situation, time of the day, etc. and have the same prefix.
      By doing so, the infrastructure provider can guarantee that no sensitive user information can leak, e.g., in a data breach, thereby complying to recent legal regulations such as the General Data Protection Regulation (GDPR)~\cite{gdpr}.

      This approach adds at most a factor of $k$ to the efficiency of range queries, since in the worst case a message to a small relevance area with one client needs to be distributed to all $k$ clients with the same prefix.
    \item
      In addition, in order to disguise origin-destination relations, users can also use shorter prefixes (i.e., coarse location information) for their starting location.
      The prefix can then be increased, e.g., when reaching higher-level parts of the road network, and be again decreased when approaching the destination, such that, e.g., no more location information is revealed after leaving the high-level road network.

      Note here that in particular hiding the destination location assumes that navigation systems are used for each ride, which we however consider a realistic assumption already in the near future to the increasing advent of self-driving cars and similar.
  \end{enumerate}
  
      We leave it as interesting open work to develop incentive models such that users do not truncate more bits than needed for $k$-anonymity, as this would decrease the information provided to the infrastructure provider, e.g., for traffic analysis and routing purposes.

  \medskip

  We would like to stress again that the three approaches presented in this section are not mutually exclusive, but can easily be combined into a single system in order to maximize the user's privacy without harming the functionality of the system.

\section{Analysis}

\subsection{Determination of Relevance Area and Message Forwarding}

For the efficiency analysis we consider the assumptions of the system model. We further assume that each client reports his full location to the routing to which it attaches.
Let $N$ be the number of clients (vehicles) and $M$ be the number of routers. 

First we consider the upper bound of distributing a message to target clients. First, the client originating a message needs to determine the set of routers. If all routers are known to the client, the client can store them e.g. in a tree-like data structure that supports efficient range queries. The Z-order has been used for range queries based on the UB-tree \ref{ramsak00integrating}. However, in the following we assume a range tree supporting fractional cascading, where vehicles are stored according to their geo-coordinates, since the Z-order is only important for assigning vehicles to routers. 
\medskip

\begin{lemma}
	A client can determine the set of routers within a rectangular relevance area in $\OO(\log M + m)$ operations, where $M$ is the overall number of routers, and $m$ the number of routers within the area.
\end{lemma}
\begin{IEEEproof}
	Follows from \cite{deberg08computational} for a two-dimensional range query, assuming that the virtual locations of the routers are known to the client and stored in a range tree supporting fractional cascading.
\end{IEEEproof}

\medskip

Afterwards, a routing request is sent to the resulting $m$ servers, either directly by the client or triggered within the router network. Each router then needs to perform a range query over its assigned clients.

\medskip

\begin{lemma} 
	A router $i$ can determine the set of the set of client within a rectangular relevance area within $\OO(\log N_i + k_i)$, where $N_i$ is the number of clients assigned to router $i$, and $k_i$ the subset of clients within the relevance area. This leads to an overall complexity of $\OO(m \log N + k)$ operations.
\end{lemma}
\begin{IEEEproof}
	The complexity for a single router follows from \cite{deberg08computational} for a two-dimensional range query, assuming that the client locations of are known to the routers and stored in a range tree supporting fractional cascading. $m$ routers perform a range query within their data structure requiring $c \log N_i + k_i$ operations for some constant $c$. This gives $\sum_{i=1}^{m} c \log N_i + \sum_{i=1}^{m} k_i = \OO (m \log N + k)$. 
\end{IEEEproof}

\medskip

After determining the clients, each router $i$ forwards the message to $k_i$ clients, leading to an overall $k = \sum_{i=1}^{m} k_i$ messages.

\medskip

\begin{corollary} Forwarding a message to k clients within a rectangular relevance area has the following complexity: 
\begin{itemize}
\item $\OO(\log M + m)$ operations of the client to determine the set of routers within the relevance area.
\item $\OO(M \log N + k)$ operations of all routers to determine the set of clients within the relevance area.
\item Communication complexity $m + k$ for $m$ routers and $k$ clients within the relevance area. \end{itemize}
\end{corollary}

\medskip

In the conventional client-server model a range query has complexity $\OO(\log N + k)$ and a communication complexity of $k$ for a single server. Once the load of maintaining client connections is distributed to $M$ servers as in our model, then the communication complexity is as follows: Assuming that clients and routers are assigned randomly, and not by location, all $M$ servers have to be requested to perform a range query. This leads to $\OO(M \log N + k)$ operations in total and a communication complexity of $M + k$. This shows that the distributed location-based assignment leads to a better performace if the set of involved servers/routers is significantly smaller than the overall set of routers (i.e. $m \ll M$).

Range queries are influenced by the variable prefix length of the client's geolocation. If clients (vehicles) opt for more location privacy, they report only a prefix, which represents an interval in the Z-order, instead of the full location. Range queries can then be performed using interval trees \cite{deberg08computational} with a similar asymptotic complexity.

\subsection{Data Structure Maintenance}

When a vehicle leaves the responsibility area of a router, then a disconnection message to the old router and a connection establishment to the new router is done, which involves only a constant number of network requests.

If a router becomes unavailable, the assigned clients will have to connect to a new server. The assignment (see \ref{fig:consistent-hashing}) affects only the current and the neighboring interval in descending Z-order. Thus the Z-order defines the fallback assignment to the next router and can be immediately used by the client.

\begin{figure}[htbp]
\centerline{\includegraphics[width=\linewidth]{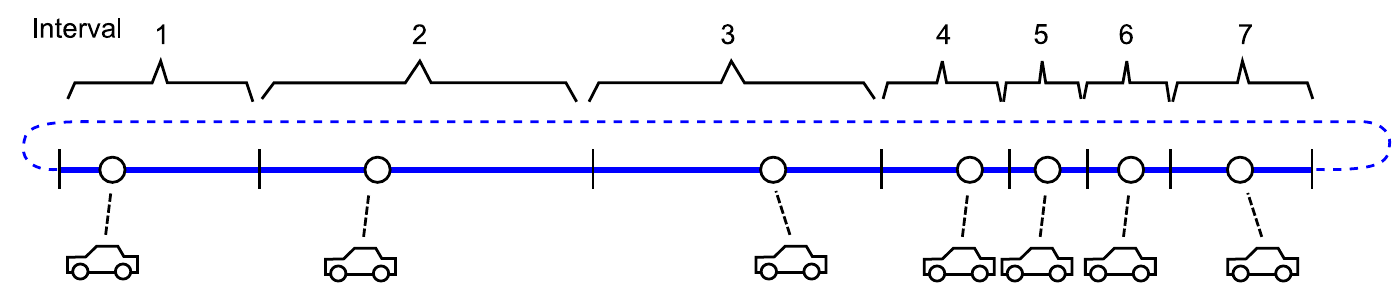}}
\caption{Intervals between vehicle locations on the Z-order curve. A distibuted hash table assigning a router to an interval number preserves a balanced assignment. }
\label{fig:density-adaption}
\end{figure}

It is important for the overall system that the load is balanced, especially when routers are added or removed. 
While in consistent hashing or in the Chord P2P network all nodes are distributed randomly on a unit interval or ring, clients are distributed regarding their real location. In order to ensure that this does not lead to impairments, routers have virtual locations that need to be assigned to locations such that a balanced assignment is guaranteed in most cases. In order to achieve this, virtual locations of the routers should follow the empirical distribution of Z-order locations of the client. This can be achieved by taking snapshots of vehicle locations in the Z-order and assigning intervals between these locations as shown in Fig.~\ref{fig:density-adaption}. Then routers can be assigned to a random interval, instead of a random location, in order to reach a balanced router-vehicle-assignment in line with the spatial distribtuion of vehicles. Using multiple virtual locations per router (as in \cite{karger97consistent}, where each server has $\log M$ representations) helps to maintain a high probability for a balanced assignment. 

\medskip

\medskip

\subsection{Alternative Routing Network} \label{sec:alternative-routing-network}

So far we assumes that routers are fully interconnected. As an alternative, routers could be organised in a P2P network. Then, clients need to request nodes within the P2P network in order to obtain routing information to one or more entry points, but they need not become member of the P2P network itself. Nodes of the P2P network need to store a routing table (routing state) that has the size of the degree of the P2P network, while the diameter of the P2P network gives the number of communication operations needed to look-up other routers. A client could maintain a link to a single routing node as an entry point to the P2P network; this has the disadvantage of revealing all location updates to this node. On the other hand a client could store the same routing table (routing state) as a node in the P2P network; this allows for faster look-ups at the cost of maintaining multiple links, depending on the size of the routing table. Table~\ref{tab:p2p-asymptotics} gives an overview of these properties, from which the routing table size (routing state) and the communication complexity of look-ups (diameter) can be derived.

\begin{table}[h]
\caption{Asymptotic properties of p2p structures over $N$ nodes}
\label{tab:p2p-asymptotics}
\footnotesize
\begin{tabular} {llll}
\hline
Name
 & \multicolumn{1}{l}{Degree} 
 & \multicolumn{1}{l}{Diameter $D$} \hfil
 & Reference
 \\
~
 & \multicolumn{1}{l}{(routing state)}
 & \multicolumn{2}{l}{(routing performance)}
 \\
\hline
\multicolumn{4}{l}{\it Structured P2P networks} \\
\quad Chord 
 & $\log_2 N$ 	
 & $\log_2 N$ 	& \scriptsize \cite{loguinov03graph_theoretic}
 \\
\quad CAN
 & 2d 	
 & $\frac12 d N^{1/d}$ 	& \scriptsize \cite{loguinov03graph_theoretic}
 \\ 
\quad Pastry 
 & $(b-1) \log_b N$ 	
 & $\log_b N$ 	& \scriptsize \cite{loguinov03graph_theoretic}
 \\
\quad Tapestry 
 & $\OO(\log_b N)$ 	
 & $\log_b N$ 	& \scriptsize \cite{lua05survey} 
  \\
\quad Kademlia 
 & $\OO(\log_b N)$ 	
 & $b \log_b N + B$ 	& \scriptsize \cite{lua05survey} 
  \\
\multicolumn{4}{l}{\it Degree-minimizing networks (structured)} \\ 
\quad Viceroy
 & $\OO(\log N)$ \textsuperscript{*}	
 & $\log N$ 	&\scriptsize  \cite{lua05survey} 
  \\  
\quad Koorde 
 & const. 	
 & $\OO(\log N)$	& \scriptsize \cite{kaashoek03koorde}
 \\
\quad Distance Halving
 & const. 	
 & $2 \log N + 3$ 	& \scriptsize \cite{mahlmann07peer} 
  \\  
\multicolumn{4}{l}{\it Further graphs} \\ 
\quad De Bruijn
 & $k$ 	
 & $\log_k N$ 	& \scriptsize \cite{loguinov03graph_theoretic}
  \\ 
\hline 
\end{tabular}

\vspace{1mm}

$b$ - alphabet size of Pastry, 
$d$ - dimension of CAN's space.

*) constant degree only if the peers are distributed perfectly \cite[p.130]{mahlmann07peer} \\
\end{table}

\medskip

\section{Conclusion and Future Work}

In this paper we have presented a combination of techniques to realise a system for message dissemination that  serves connected vehicles without centralised control and without tracking location information. We could show by formal arguments that there are no performance losses in terms of rang query complexity and communication complexity in comparison to the conventional client-server-model.

There are certain interesting issues for future work: 
In our model, location information can be fine grained or coarse grained, leaving a trade-off between efficiency and privacy. It is an open issue how to derive concrete parameters for maintaining k-anonymity. Furthermore, it is an open issue to develop incentive models such that users do not hide more location data than needed for k-anonymity, as this would decrease the efficiency of message routing and its usefulness for road traffic analysis.

Furthermore, mobility follows regular daily and seasonal patters, which could be exploited for an optimal choice of the number of routers, the organisation of the routing nodes and the parameterization of the distributed hash function that assigns routers with the respective vehicles.



\vspace{12pt}

\end{document}